\documentclass[12pt]{article}
\usepackage{graphicx}
\usepackage{lscape}
\usepackage{citesort}

\begin{document}

\def\ds{\displaystyle}
\def\beq{\begin{equation}}
\def\eeq{\end{equation}}
\def\bea{\begin{eqnarray}}
\def\eea{\end{eqnarray}}
\def\beeq{\begin{eqnarray}}
\def\eeeq{\end{eqnarray}}
\def\ve{\vert}
\def\vel{\left|}
\def\ver{\right|}
\def\nnb{\nonumber}
\def\ga{\left(}
\def\dr{\right)}
\def\aga{\left\{}
\def\adr{\right\}}
\def\lla{\left<}
\def\rra{\right>}
\def\rar{\rightarrow}
\def\nnb{\nonumber}
\def\la{\langle}
\def\ra{\rangle}
\def\ba{\begin{array}}
\def\ea{\end{array}}
\def\tr{\mbox{Tr}}
\def\ssp{{\Sigma^{*+}}}
\def\sso{{\Sigma^{*0}}}
\def\ssm{{\Sigma^{*-}}}
\def\xis0{{\Xi^{*0}}}
\def\xism{{\Xi^{*-}}}
\def\qs{\la \bar s s \ra}
\def\qu{\la \bar u u \ra}
\def\qd{\la \bar d d \ra}
\def\qq{\la \bar q q \ra}
\def\gGgG{\la g^2 G^2 \ra}
\def\q{\gamma_5 \not\!q}
\def\x{\gamma_5 \not\!x}
\def\g5{\gamma_5}
\def\sb{S_Q^{cf}}
\def\sd{S_d^{be}}
\def\su{S_u^{ad}}
\def\ss{S_s^{??}}
\def\sbp{{S}_Q^{'cf}}
\def\sdp{{S}_d^{'be}}
\def\sup{{S}_u^{'ad}}
\def\ssp{{S}_s^{'??}}
\def\sig{\sigma_{\mu \nu} \gamma_5 p^\mu q^\nu}
\def\fo{f_0(\frac{s_0}{M^2})}
\def\ffi{f_1(\frac{s_0}{M^2})}
\def\fii{f_2(\frac{s_0}{M^2})}
\def\O{{\cal O}}
\def\sl{{\Sigma^0 \Lambda}}
\def\es{\!\!\! &=& \!\!\!}
\def\ap{\!\!\! &\approx& \!\!\!}
\def\ar{&+& \!\!\!}
\def\ek{&-& \!\!\!}
\def\kek{\!\!\!&-& \!\!\!}
\def\cp{&\times& \!\!\!}
\def\se{\!\!\! &\simeq& \!\!\!}
\def\eqv{&\equiv& \!\!\!}
\def\kpm{&\pm& \!\!\!}
\def\kmp{&\mp& \!\!\!}


\def\simlt{\stackrel{<}{{}_\sim}}
\def\simgt{\stackrel{>}{{}_\sim}}


\title{
         {\Large
                 {\bf
Analysis of  $\Lambda_b \rar \Lambda \ell^+ \ell^-$ decay within family
non--universal $Z^\prime$ model 
                 }
         }
      }

\author{\vspace{1cm}\\
{\small T. M. Aliev \thanks {e-mail:
taliev@metu.edu.tr}~\footnote{permanent address:Institute of
Physics,Baku,Azerbaijan}\,\,, M. Savc{\i} \thanks
{e-mail: savci@metu.edu.tr}} \\
{\small Physics Department, Middle East Technical University,
06531 Ankara, Turkey }\\
       }

\date{}

\begin{titlepage}
\maketitle
\thispagestyle{empty}

\begin{abstract}
We perform a comprehensive analysis of the rare  $\Lambda_b \rar \Lambda
\ell^+ \ell^-$ decay in the framework of family non--universal $Z^\prime$
model. It is shown that $Z^\prime$ gives considerable contribution to the
decay width. Zero positions of the forward--backward asymmetry and
$\alpha_\theta$ parameter are shifted to the left compared to the Standard
Model result. The obtained results could be tested in near future at LHC--b.   
\end{abstract}

~~~PACS numbers: 12.15.Ji, 12.60.--i, 13.30.--a, 13.88.+e
\end{titlepage}

\section{Introduction}
Recently, CDF Collaboration has reported the first observation of the
baryonic flavor--changing neutral current (FCNC) decay $\Lambda_b \rar \Lambda
\mu^+ \mu^-$ \cite{R11101}. At quark level this decay is described by the $b
\rar s \mu^+ \mu^-$ transition, which is forbidden in SM at tree level and
take place only at loop level. Therefore this transition represents an
excellent channel in searching new physics beyond the Standard Model (SM). 

The new physics effects in rare decays manifest themselves either through
the Wilson coefficients which are different compared to the one in the SM
counterpart, or via the new operator structures in an effective Hamiltonian
which are absent in the SM.

In this regard, the study of the baryonic flavor changing neutral currents
is quite promising, since they could maintain the helicity structure of the
effective Hamiltonian, in contrary to the mesonic case where it
is lost through hadronization \cite{R11102}. Following the observation of
rare $\Lambda_b \rar \Lambda \ell^+ \ell^-$ decay by CDF Collaboration, the
next step is a comprehensive study of various weak, electromagnetic and
strong decays of the $\Lambda_b$ baryons. Note that $\Lambda_b \rar \Lambda
\ell^+ \ell^-$ decay is planned to be investigated in detail at Large Hadron
Collider (LHC). Having the present experimental motivation, it is timely now
to study the properties of the heavy baryons theoretically.

Rare baryonic $\Lambda_b \rar \Lambda \ell^+ \ell^-$ decay has been
investigated in the SM in numerous works (see for example \cite{R11103} and
references therein). The branching ratio of the $\Lambda_b \rar \Lambda
\mu^+ \mu^-$ decay is found to have the value $Br(\Lambda_b \rar \Lambda
\mu^+ \mu^-) = (4.0 \pm 1.2) \times 10^{-6}$. Many physical observables
such as the branching ratio, forward--backward asymmetry ${\cal A}_{FB}$,
lepton polarization induced by the $b \to s \ell^+ \ell^-$ transition are
very sensitive to the existence of new physics. In addition to these
observables in the $\Lambda_b \rar \Lambda \ell^+ \ell^-$ decay, measurement
of the of the polarizations $\Lambda_b$ and $\Lambda$ are very useful in
this respect. 

In the present work, we study the rare baryonic $\Lambda_b \rar \Lambda
\ell^+ \ell^-$ decay within non--universal $Z^\prime$ model. The
non--universal $Z^\prime$ models appear in certain string construction
\cite{R11104} and $E_6$ models \cite{R11105} by introducing family
non--universal $U(1)$ gauge symmetry. The family non--universal  $Z^\prime$
model has comprehensively been developed in \cite{R11106}. 

The possible manifestation of non--universal  $Z^\prime$ bosons in various
$B$ meson sectors
has been investigated in detail in many works \cite{R11107,R11108,R11109}
(for a recent review, see \cite{R11110}).

The plan of the work is as follows. In section 2, the effective Hamiltonian
describing $b \to s \ell^+ \ell^-$ transition is presented in both standard and
$Z^\prime$ models. Using this effective Hamiltonian, the matrix
element for the $\Lambda_b \rar \Lambda \ell^+ \ell^-$ decay is then
obtained. In this section we also present the expressions of various
physical observables in $Z^\prime$ model. Section 3 is devoted to the
numerical analysis of the obtained physical observables. We present our
conclusions in section 3.

\section{ $\Lambda_b \rar \Lambda \ell^+ \ell^-$ decay in the SM and
family non--universal $Z^\prime$ model}

At quark level, $\Lambda_b \rar \Lambda \ell^+ \ell^-$ decay is described in
the SM by the $b \to s \ell^+ \ell^-$ transition. The effective
Hamiltonian for this transition in the SM is \cite{R11111,R11112}
\bea
\label{e11101}
{\cal H} = - {4G \over \sqrt{2}} V_{tb} V_{ts}^\ast \sum_{i=1}^{10} C_i
{\cal O}_i ~,
\eea
where ${\cal O}_i(\mu)$ are the local operators and $C_i$ are the
corresponding Wilson coefficients.
The expressions of all local operators can be found in \cite{R11111,R11112}.
Here, the terms proportional to $V_{ub} V_{us}^\ast$ have been neglected,
since $\vel V_{ub} V_{us}^\ast/V_{tb} V_{ts}^\ast \ver \le 0.02$.
In further discussion we need the operators ${\cal O}_7$, ${\cal O}_9$ and
${\cal O}_{10}$, whose expressions are given as follows:
\bea
\label{e11102} 
{\cal O}_7 \es {e \over g_s^2} m_b \bar{s} \sigma_{\mu\nu} P_R b \bar{\ell}
\gamma_\mu \ell~, \nnb \\
{\cal O}_9 \es {e^2 \over g_s^2} \bar{s} \gamma_\mu P_L b \bar{\ell} 
\gamma_\mu \ell~, \nnb \\
{\cal O}_{10} \es {e^2 \over g_s^2} \bar{s} \gamma_\mu P_L b \bar{\ell}                
\gamma_\mu \gamma_5 \ell~,
\eea
where $P_{L,R} = (1 \mp \gamma_5)/2$.
The values of the Wilson coefficients at $\mu=m_b$ scale at next--to--next
leading logarithm (NNLL) are calculated in many works (for example see
\cite{R11113} and references therein).

This effective Hamiltonian leads to the following result for the matrix
element of the $b \to s \ell^+ \ell^-$ transition,
\bea
\label{e11103}
{\cal M} = {G \over \sqrt{2}} V_{tb} V_{ts}^\ast \Bigg\{C_9^{eff}
\bar{s} \gamma_\mu P_L b \bar{\ell} \gamma_\mu \ell +
C_{10} \bar{s} \gamma_\mu P_L b \bar{\ell} \gamma_\mu \gamma_5 \ell
\ek 2 C_7^{eff} {1\over q^2} \bar{s} \gamma_\mu P_R b \bar{\ell} \gamma_\mu
\ell \Bigg\}~.
\eea
The Wilson coefficient $C_9^{eff}$ contains short and long distance
contributions whose expression is given as,
\bea
\label{e11104}
C_9^{eff} = {4 \pi \over \alpha_s} C_9 Y_{SD}(z,\hat{s}) +
Y_{LD}(z,\hat{s})~. \nnb
\eea
In this expression $z=m_c/m_b$, $hat{s} = q^2/m_b^2$. The term
$Y_{SD}(z,\hat{s})$ represents the contributions coming from local
four--quark operators. The Wilson
coefficient $C_9$ receives also long distance contributions $Y_{LD}$ due to
the real $\bar{c}c$ intermediate states, i.e., $J/\psi$, $\psi^\prime$, etc.
Explicit expression of $Y_{LD}(z,\hat{s})$ and detailed discussion about it
can be found in \cite{R11114}.

Let us now discuss how non--universal $Z^\prime$ effects modify the
effective Hamiltonian. For this aim we will follow the model presented in
\cite{R11106}. In this model interactions of $Z^\prime$ with the
right--handed quarks are flavor diagonal. The effective Hamiltonian for the
$b \rar s \ell^+ \ell^-$ transition in the presence of $Z^\prime$ is
modified as \cite{R11107}
\bea
\label{e11105}
{\cal H}_{eff}^{Z^\prime}(b\to s \ell^+ \ell^-) \es - {2 G_F \over
\sqrt{2}} V_{tb} V_{ts}^\ast \Bigg[ - {B_{sb}^L B_{\ell\ell}^L \over V_{tb}
V_{ts}^\ast } (\bar{s}b)_{V-A} (\ell\ell)_{V-A} \nnb \\
\ek {B_{sb}^L B_{\ell\ell}^R \over V_{tb} V_{ts}^\ast } (\bar{s}b)_{V-A}
(\ell\ell)_{V+A} \Bigg] + h.c.~,
\eea
where $B_{sb}^L$ and $B_{\ell\ell}^{L,R}$ correspond to the chiral
$Z^\prime$ couplings with quarks and leptons.

Assuming that there is no considerable running effects between
$m_{Z^\prime}$ and $m_W$ scales, $Z^\prime$ contribution leads to the
modification of the Wilson coefficients, i.e., $C_{9,10}^\prime (m_W) =
C_{9,10}^{SM}(m_W) + \Delta C_{9,10} (m_W)$. In other words, the
$Z^\prime$ part of the effective Hamiltonian for the $b \to s \ell^+ \ell^-$
transition can be written as
\bea
\label{e11106}
{\cal H}^{Z^\prime} = - {4 G_F \over \sqrt{2}} V_{tb} V_{ts}^\ast \Big[
\Delta C_9 {\cal O}_9 + \Delta C_{10} {\cal O}_{10} \Big] + h.c.~,
\eea
where
\bea
\label{nolabel}
\Delta C_9^{Z^\prime} \es - {g_s^2 \over e^2} {B_{sb}^L \over V_{tb} V_{ts}^\ast} 
\left(B_{\ell\ell}^L + B_{\ell\ell}^R \right)~, \nnb \\
\Delta C_{10}^{Z^\prime} \es {g_s^2 \over e^2} {B_{sb}^L \over V_{tb} V_{ts}^\ast} 
\left(B_{\ell\ell}^L - B_{\ell\ell}^R \right)~. \nnb
\eea
From $m_W$ to $m_b$ scale the running effects are the same as in SM
\cite{R11115}. In further numerical analysis we shall use
\bea
\label{e11107}
C_9^\prime (m_b) \es 0.0682 -28.82 {B_{sb}^L \over V_{tb} V_{ts}^\ast}
\left(B_{\ell\ell}^L + B_{\ell\ell}^R \right)~, \nnb \\
C_{10}^\prime (m_b) \es -0.0695 + 28.82 {B_{sb}^L \over V_{tb} V_{ts}^\ast}
\left(B_{\ell\ell}^L - B_{\ell\ell}^R \right)~.
\eea

In result, in order to implement the effects coming from $Z^\prime$ boson it is
enough to make the replacements
\bea
\label{e11108}
C_9 (m_b)^{SM} &\to& C_9^\prime (m_b)~, \nnb \\
C_{10}^{SM} (m_b) &\to& C_{10}^\prime (m_b).  
\eea
Hence, in the considered version of flavor non--universal $Z^\prime$ model
there does not appear any new operator structure other than those in SM.

As a result, including the $Z^\prime$ contribution
by making the replacements given in Eq. (\ref{e11108}),
the matrix element responsible for $b \to s \ell^+ \ell^-$
transition coincides with Eq. (\ref{e11103}), i.e., 
\bea
\label{e11109}
{\cal M} \es {G\alpha \over 2 \sqrt{2} \pi} V_{tb} V_{ts}^\ast \Bigg\{
C_9^\prime \bar{s} \gamma_\mu (1-\gamma_5) b \bar{\ell} \gamma_\mu \ell +
C_{10}^\prime \bar{s} \gamma_\mu (1-\gamma_5) b \bar{\ell} 
\gamma_\mu \gamma_5 \ell \nnb \\
\ek 2 m_b C_7 {1\over q^2} \bar{s} i \sigma_{\mu\nu} q^\nu (1+\gamma_5) b
\bar{\ell}\gamma_\mu \ell \Bigg\}~.
\eea

The amplitude of exclusive $\Lambda_b \rar \Lambda \ell^+ \ell^-$ decay,
which is described at quark level by the $b \to s \ell^+ \ell^-$ transition
can be obtained from Eq. (\ref{e11109}) by replacing it between the initial
and final baryon states. These matrix elements are parametrized in terms of
the form factors as follows:
\bea
\label{e11110}
\lla \Lambda \vel \bar{s} \gamma_\mu (1 - \gamma_5) b \ver \Lambda_b \rra \es 
\bar{u}_\Lambda (p) \Big[ \gamma_\mu f_1(q^2) + i \sigma_{\mu\nu} q^\nu
f_2(q^2) + q_\mu f_3(q^2) \nnb \\
\ek \gamma_\mu \gamma_5 g_1(q^2) - i \sigma_{\mu\nu} \gamma_5 q^\nu g_2(q^2) 
- q_\mu \gamma_5 g_3(q^2) \Big] u_{\Lambda_b}(p_b)~, \\ \nnb \\
\label{e11111}
\lla \Lambda \vel \bar{s} i \sigma_{\mu\nu} q^\nu (1 + \gamma_5) b \ver
\Lambda_b(p_b) \rra \es 
\bar{u}_\Lambda(p) \Big[ \gamma_\mu f_1^T(q^2) + 
i \sigma_{\mu\nu} q^\nu f_2^T(q^2) + q_\mu f_3^T(q^2) q_\mu \nnb \\
\ar \gamma_\mu \gamma_5 g_1^T(q^2) + i \sigma_{\mu\nu} 
\gamma_5 q^\nu g_2^T(q^2) + q_\mu \gamma_5 g_3^T(q^2) \Big] u_{\Lambda_b} (p_b)~.
\eea

The matrix element for exclusive $\Lambda_b \rar \Lambda \ell^+ \ell^-$
decay can easily be obtained in terms of twelve form factors from Eqs.
(\ref{e11109}--\ref{e11110}), and we find that
\bea
\label{e11112}
{\cal M} \es {G \alpha \over 8 \sqrt{2}\pi} V_{tb}V_{ts}^\ast \Bigg\{
\bar{\ell} \gamma^\mu (1-\gamma_5)  \ell \, \bar{u}_\Lambda(p) \Big\{
(A_1-D_1) \gamma_\mu (1+\gamma_5) +
(B_1+E_1) \gamma_\mu (1-\gamma_5) \nnb \\
\ar i \sigma_{\mu\nu} q^\nu \big[ (A_2-D_2) (1+\gamma_5) + 
(B_2-E_2) (1-\gamma_5) \big] \nnb \\
\ar q_\mu \big[ (A_3-D_3) (1+\gamma_5) + (B_3-E_3) (1-\gamma_5) \big]\Big\}
u_{\Lambda_b}(p_b) \nnb \\
\ar \bar{\ell} \gamma^\mu (1+\gamma_5)  \ell \, \bar{u}_\Lambda(p) \Big\{
(A_1+D_1) \gamma_\mu (1+\gamma_5) +
(B_1+E_1) \gamma_\mu (1-\gamma_5) \nnb \\
\ar i \sigma_{\mu\nu} q^\nu \big[ (A_2+D_2) (1+\gamma_5) +           
(B_2+E_2) (1-\gamma_5) \big] \nnb \\
\ar q_\mu \big[ (A_3+D_3) (1+\gamma_5) + (B_3+E_3) (1-\gamma_5) \big]\Big\}
u_{\Lambda_b}(p_b) \Bigg\}~,
\eea
where
\bea
\label{nolabel}
A_1 \es - {2 m_b\over q^2} C_7 \ga f_1^T + g_1^T \dr + 
C_9 \ga f_1-g_1 \dr ~,\nnb \\
A_2 \es A_1 \ga 1 \rar 2 \dr ~,\nnb \\
A_3 \es A_1 \ga 1 \rar 3 \dr ~,\nnb \\
B_i \es A_i \ga g_i \rar - g_i;~g_i^T \rar - g_i^T \dr ~,\nnb \\
D_1 \es C_{10} \ga f_1-g_1 \dr ~,\nnb \\
D_2 \es D_1 \ga 1 \rar 2 \dr ~, \nnb \\
D_3 \es D_1 \ga 1 \rar 3 \dr ~,\nnb \\
E_i \es D_i \ga g_i \rar - g_i \dr ~.\nnb
\eea

Adopting the same convention and notation as in \cite{R11116}, the helicity
amplitudes are given by the following expressions:
\bea
\label{e11113}
{\cal M}_{+1/2}^{++} \es 2 m_\ell \sin\theta \Big( H_{+1/2,+1}^{(1)} +
H_{+1/2,+1}^{(2)}\Big) + 2 m_\ell \cos\theta \Big( H_{+1/2,0}^{(1)} +
H_{+1/2,0}^{(2)}\Big) \nnb \\
\ar 2 m_\ell \Big( H_{+1/2,t}^{(1)} -
H_{+1/2,t}^{(2)}\Big)~, \nnb \\
{\cal M}_{+1/2}^{+-} \es - \sqrt{q^2} (1-\cos\theta) \Big[
(1-v) H_{+1/2,+1}^{(1)} + (1+v) H_{+1/2,+1}^{(2)}\Big] -
\sqrt{q^2} \sin\theta  \Big[(1-v) H_{+1/2,0}^{(1)} \nnb \\
\ar (1+v) H_{+1/2,0}^{(2)}\Big]~, \nnb \\
{\cal M}_{+1/2}^{-+} \es \sqrt{q^2} (1+\cos\theta) \Big[
(1+v) H_{+1/2,+1}^{(1)} + (1-v) H_{+1/2,+1}^{(2)}\Big] -
\sqrt{q^2} \sin\theta  \Big[(1+v) H_{+1/2,0}^{(1)} \nnb \\
\ar (1-v) H_{+1/2,0}^{(2)}\Big]~, \nnb \\
{\cal M}_{+1/2}^{--} \es - 2 m_\ell \sin\theta \Big(
H_{+1/2,+1}^{(1)} +
H_{+1/2,+1}^{(2)}\Big) - 2 m_\ell \cos\theta \Big( H_{+1/2,0}^{(1)} +
H_{+1/2,0}^{(2)}\Big) \nnb \\
\ar 2 m_\ell \Big( H_{+1/2,t}^{(1)} -
H_{+1/2,t}^{(2)}\Big)~, \nnb \\
{\cal M}_{-1/2}^{++} \es - 2 m_\ell \sin\theta \Big(
H_{-1/2,-1}^{(1)} +
H_{-1/2,-1}^{(2)}\Big) + 2 m_\ell \cos\theta \Big( H_{-1/2,0}^{(1)} +
H_{-1/2,0}^{(2)}\Big) \nnb \\
\ar 2 m_\ell \Big( H_{-1/2,t}^{(1)} -
H_{-1/2,t}^{(2)}\Big)~, \nnb \\
{\cal M}_{-1/2}^{+-} \es - \sqrt{q^2} (1+\cos\theta) \Big[
(1-v) H_{-1/2,-1}^{(1)} + (1+v) H_{-1/2,-1}^{(2)}\Big] -
\sqrt{q^2} \sin\theta  \Big[(1-v) H_{-1/2,0}^{(1)} \nnb \\
\ar (1+v) H_{-1/2,0}^{(2)}\Big]~, \nnb \\
{\cal M}_{-1/2}^{-+} \es \sqrt{q^2} (1-\cos\theta) \Big[
(1+v) H_{-1/2,-1}^{(1)} + (1-v) H_{-1/2,-1}^{(2)}\Big] -
\sqrt{q^2} \sin\theta  \Big[(1+v) H_{-1/2,0}^{(1)} \nnb \\
\ar (1-v) H_{-1/2,0}^{(2)}\Big]~, \nnb \\
{\cal M}_{-1/2}^{--} \es 2 m_\ell \sin\theta \Big( H_{-1/2,-1}^{(1)} +
H_{-1/2,-1}^{(2)}\Big) - 2 m_\ell \cos\theta \Big( H_{-1/2,0}^{(1)} +
H_{-1/2,0}^{(2)}\Big) \nnb \\
\ar 2 m_\ell \Big( H_{-1/2,t}^{(1)} -
H_{-1/2,t}^{(2)}\Big)~.
\eea
Here
\bea      
\label{e11114}
H_{\pm 1/2,\pm1}^{(1)}   \es H_{1/2,1}^{(1)\,V}   \pm H_{1/2,1}^{(1)\,A}~,   \nnb \\
H_{\pm 1/2,\pm1}^{(2)}   \es H_{1/2,1}^{(2)\,V}   \pm H_{1/2,1}^{(2)\,A}~,   \nnb \\
H_{\pm 1/2,0}^{(1,2)}    \es H_{1/2,0}^{(1,2)\,V} \pm H_{1/2,1}^{(1,2)\,A}~, \nnb \\
H_{\pm 1/2,t}^{(1,2)}    \es H_{1/2,t}^{(1,2)\,V} \pm H_{1/2,t}^{(1,2)\,A}~,
\eea
where $\theta$ is the angle of the positron in the rest frame of the
intermediate boson with respect to its helicity axes.
The superscripts in ${\cal M}$ correspond to the helicities of leptons and
subscript denotes the helicity of the $\Lambda$ baryon. The amplitudes
$H_{\lambda,\lambda_W}^{V,A}$ are defined as:
\bea
\label{e11115}
H_{1/2,1}^{(1)\,V} \es - \sqrt{Q_-} \Big[ F_1^V - (m_{\Lambda_b}+m_\Lambda) F_2^V
\Big]~, \nnb \\
H_{1/2,1}^{(1)\,A} \es - \sqrt{Q_+} \Big[ F_1^A + (m_{\Lambda_b}-m_\Lambda) F_2^A
\Big]~, \nnb \\
H_{1/2,1}^{(2)\,V} \es H_{1/2,1}^{(1)\,V} (F_1^V \rar F_3^V,~F_2^V \rar
F_4^V)~, \nnb \\
H_{1/2,1}^{(2)\,A} \es H_{1/2,1}^{(1)\,A} (F_1^A \rar F_3^A,~F_2^A \rar
F_4^A)~, \nnb \\
H_{1/2,0}^{(1)\,V} \es - \frac{1}{\sqrt{q^2}} \Big\{ \sqrt{Q_-} 
\Big[ (m_{\Lambda_b}+m_\Lambda) F_1^V - q^2 F_2^V \Big]
\Big\}~,\nnb \\
H_{1/2,0}^{(1)\,A} \es - \frac{1}{\sqrt{q^2}} \Big\{ \sqrt{Q_+} 
\Big[ (m_{\Lambda_b}-m_\Lambda) F_1^A + q^2 F_2^A \Big]
\Big\}~,\nnb \\
H_{1/2,0}^{(2)\,V} \es H_{1/2,0}^{(1)\,V} (F_1^V \rar F_3^V,~F_2^V \rar
F_4^V) ~,\nnb \\
H_{1/2,0}^{(2)\,A} \es H_{1/2,0}^{(1)\,A} (F_1^A \rar F_3^A,~F_2^A \rar
F_4^A) ~,\nnb \\
H_{1/2,t}^{(1)\,V} \es - \frac{1}{\sqrt{q^2}} \Big\{ \sqrt{Q_+} 
\Big[ (m_{\Lambda_b}-m_\Lambda) F_1^V + q^2 F_5^V \Big]
\Big\}~,\nnb \\
H_{1/2,t}^{(1)\,A} \es - \frac{1}{\sqrt{q^2}} \Big\{ \sqrt{Q_-} 
\Big[ (m_{\Lambda_b}+m_\Lambda) F_1^A - q^2 F_5^A \Big]
\Big\}~,\nnb \\
H_{1/2,t}^{(2)\,V} \es H_{1/2,t}^{(1)\,V} (F_1^V \rar F_3^V,~F_5^V \rar 
F_6^V) ~,\nnb \\
H_{1/2,t}^{(2)\,A} \es H_{1/2,t}^{(1)\,A} (F_1^A \rar F_3^A,~F_5^A \rar
F_6^A) ~,
\eea
where
\bea
Q_+ \es (m_{\Lambda_b}+m_\Lambda)^2 - q^2~,\nnb \\
Q_- \es (m_{\Lambda_b}-m_\Lambda)^2 - q^2~,\nnb
\eea 
and
\bea
\label{e11116}
F_1^V \es  A_1-D_1+B_1-E_1~,\nnb \\
F_1^A \es  A_1-D_1-B_1+E_1~,\nnb \\
F_2^V \es  F_1^V (1\rar 2)~,\nnb \\
F_2^A \es  F_1^A (1\rar 2)~,\nnb \\
F_3^V \es  A_1+D_1+B_1+E_1~,\nnb \\
F_3^A \es  A_1+D_1-B_1-E_1~,\nnb \\
F_4^V \es  F_3^V (1\rar 2)~,\nnb \\
F_4^A \es  F_3^A (1\rar 2)~,\nnb \\
F_5^V \es  F_1^V (1\rar 3)~,\nnb \\
F_5^A \es  F_1^A (1\rar 3)~,\nnb \\
F_6^V \es  F_3^V (1\rar 3)~,\nnb \\
F_6^A \es  F_3^A (1\rar 3)~.
\eea
Rest of the helicity amplitudes entering into Eq. (\ref{e11113}) can be obtained 
from the following relations,
\bea
\label{e11117}
H_{-\lambda,-\lambda_W}^{V,(A)} = +(-) H_{\lambda,\lambda_W}^{V,(A)}~.
\eea
The square of the matrix element for the $\Lambda_b \rar\Lambda \ell^+\ell^-$ 
decay is given as
\bea
\label{e11118}
\vel {\cal M} \ver^2 \es \vel {\cal M}_{+1/2}^{++} \ver^2 +
\vel {\cal M}_{+1/2}^{+-} \ver^2 + \vel {\cal M}_{+1/2}^{-+} \ver^2 + 
\vel {\cal M}_{+1/2}^{--} \ver^2 \nnb \\
\ar \vel {\cal M}_{-1/2}^{++} \ver^2 +
\vel {\cal M}_{-1/2}^{+-} \ver^2 + \vel {\cal M}_{-1/2}^{-+} \ver^2
+ \vel {\cal M}_{-1/2}^{--} \ver^2~.
\eea

In further discussions, we shall study the following observables:

1) $q^2$ dependence of the differential branching ratio $d {\cal B}/
dq^2$. The expression of the differential branching ratio
can be obtained by integrating the double differential branching ratio
over $x=\cos\theta$ whose explicit form is presented in the appendix. 

2)  Forward--backward asymmetry,
\bea
\label{e111??}
{\cal A}_{FB} = \frac{\ds{\int_0^1\frac{d\Gamma}{dsdx}}\,dx -
\ds{\int_{-1}^0\frac{d\Gamma}{dsdx}}\,dx}
{\ds{\int_0^1\frac{d\Gamma}{dsdx}}\,dx +
\ds{\int_{-1}^0\frac{d\Gamma}{dsdx}}\,dx}~, \nnb
\eea
and again, its explicit form can easily be obtained from $d\Gamma/dq^2 dx$.

3) The polar angle $\theta_\Lambda$ distribution of the cascade $\Lambda \to
a + b$ decay. This distribution is given by
\bea
\label{e111??}
\frac{d\Gamma}{dq^2\,d\!\cos\theta_\Lambda} \sim 1 + \alpha \alpha_\Lambda
\cos\theta_\Lambda~. \nnb
\eea

4) The polar angle distribution of the cascade $V^\ast \to \ell^+ \ell^-$
decay, whose explicit form is given as

\bea
\label{e111??}
\frac{d\Gamma}{dq^2\,d\!\cos\theta} \sim 1 + 2 \alpha_\theta \cos\theta +
\beta_\theta \cos^2\theta~. \nnb
\eea

5) The polarization asymmetry parameter $\alpha_{\Lambda_b}$, when the
polarization of the initial $\Lambda_b$ is considered. The
parameter $\alpha_{\Lambda_b}$ has the following form

\bea
\label{e111??}
\frac{\ds d\Gamma}{\ds dq^2\,d\!\cos\theta_{\Lambda_b}} \sim 1 -
\alpha_{\Lambda_b} {\cal P} \cos\theta_{\Lambda_b}~, \nnb
\eea
where ${\cal P}$ is the polarization of $\Lambda_b$ and
$\theta_{\Lambda_b}$ is the angle between the polarization of $\Lambda_b$ 
with its momentum.

\section{Numerical analysis}

In this section we present our numerical calculations of the physical
observables given in the previous section, in family non--universal
$Z^\prime$ model. As has already been mentioned, in this considered version
version of the family non--universal $Z^\prime$ model no new operators
appear compared to the SM, and hence the effect of $Z^\prime$ contribution
is implemented by making modifications in the new Wilson coefficients $C_9$
and $C_{10}$. These modifications are described by four new parameters $\vel
B_{sb}^L \ver$, $\varphi_s^L$, $B_{\ell\ell}^L$ and $B_{\ell\ell}^R$.
Constraints to  $\vel B_{sb}^L \ver$ and $\varphi_s^L$ coming from $\bar{B}_s
-B_s$ mixing, $B \to \pi K^{(\ast)},\rho  K^{(\ast)}$ are studied in
detail in \cite{R11109}. Moreover, in these work, restrictions to the
parameters $B_{\ell\ell}^L$ and $B_{\ell\ell}^R$ that come from $B \to X_s \mu^+
\mu^-$, $B \to K^{(\ast)} \mu^+ \mu^-$ and $B_s \to \mu^+ \mu^-$ decays are
obtained. Recently, more new data on the above--mentioned decays
have been accumulated in Tevatron and LHC, and therefore some of the
constraints on these parameters might might be changed. In Table 1 we
present the numerical results of these parameters, where S1 and S2
correspond to UT--fit collaboration's two fitting results for the
$\bar{B}_s-B_s$ mixing \cite{R11117}.

\begin{table}[h]    
\renewcommand{\arraystretch}{1.5}
\addtolength{\arraycolsep}{3pt}
$$
\begin{array}{|l|c|c|c|c|}
\hline
 & \vel B_{sb}^L \ver\times 10^{-3} & \varphi_s^{L[0]} & 
B_{\mu\mu}^L \times 10^{-2} & B_{\mu\mu}^R\times 10^{-2} \\ \hline
S1 & 1.09 \pm 0.22 &^ -72 \pm 7 & -4.75 \pm 2.44 & 1.97 \pm 2.24  \\ \hline
S2 & 2.20 \pm 0.15 &^ -82 \pm 4 & -1.83 \pm 0.82 & 0.68 \pm 0.85  \\ \hline
\end{array}
$$
\caption{The values of the input parameters for the $Z^\prime$ couplings.}
\renewcommand{\arraystretch}{1}
\addtolength{\arraycolsep}{-3pt}
\end{table}

In order to maximize the effects of $Z^\prime$, we choose the maximum values
of these parameters. In the case of S1, we use $B_{sb}^L = 1.31 \times
10^{-3}$, $\varphi_s^L=-79^0$, $B_{\ell\ell}^L + B_{\ell\ell}^R = -6.7\times
10^{-2}$, $B_{\ell\ell}^L - B_{\ell\ell}^R = -9.3\times 10^{-2}$.

Other input parameters that are essential in performing numerical analysis
are the form factors. All form factors responsible for the $\Lambda_b \to
\Lambda$ transition within the light cone QCD sum rules are calculated in
\cite{R11103} and these results have been used in the present numerical
analysis.

In Figs. 1--4, we present the $q^2$ dependence of the differential branching
ratio, forward--backward asymmetry, the polar angle $\theta_{\Lambda}$
distribution of the cascade $\Lambda \to a + b$ decay, the polar angle
$\alpha_\theta$ distribution of the cascade $V^\ast \to \ell^+ \ell^-$ decay,
respectively.

From these figures we get the following results. 

\begin{itemize}

\item The effect of S1  set of parameters to the differential branching
ratio is larger compared to the S2 case. Branching ratio in both cases
exceeds the SM prediction. For example, in the ``low" $q^2$ region
branching ratio is enhanced about 100\% in S1 and 40\% in S2 cases for
the central values of the parameters, compared to the SM predictions.

\item Zero position of the forward--backward asymmetry is shifted to left
compared to the SM case. Therefore, experimental determination of the zero
position is quite important for establishing new physics beyond the SM. 

\item In contrast to the differential branching ratio $d{\cal B}/dq^2$ and
forward--backward asymmetry ${\cal A}_{AB}$, the parameter $\alpha_\Lambda$
does practically coincide with the $Z^\prime$ model prediction for both S1
and S2 set of parameters.

\item Similar to the forward--backward asymmetry case, the zero position of
the asymmetry parameter $\alpha_\theta$ is very sensitive to the new physics
effects and it is shifted to the left compared to the SM prediction.

\item The parameters $\alpha_{\Lambda_b}$ and $\beta_\theta$ 
are not sensitive to the new physics effects.

\end{itemize}

As a result, the physical observables $d{\cal B}/dq^2$, ${\cal A}_{FB}$ and
$\alpha_\theta$ are very sensitive to the new physics contributions. All
these results can be checked in new future at LHCb.

Our final remark is as follows. Investigation of the lepton polarizations is
quite an efficient tool for searching new physics. There immediately follows
then the following question: how sensitive lepton polarizations are to the
new flavor--non diagonal $Z^\prime$ effects. We are planning to discuss this
issue elsewhere in future. 

In conclusion, the effects of new family non--universal $Z^\prime$ model 
contributions have
been studied for the $\Lambda_b \rar \Lambda \ell^+ \ell^-$ decay. We found
that the contributions of the family non--universal $Z^\prime$ model to the
differential branching ratio is quite significant and enhances the SM
predictions considerably. We also observe that the analysis of the zero
positions of the forward--backward asymmetry and parameter $\alpha_\theta$
are both shifted to the left compared to the SM prediction and can serve as
a very efficient tool for establishing new physics beyond the SM.

\section*{Acknowledgments}
We thank S. R. Choudhury and N. Gaur for their collaboration at the early
stage of this work.

\newpage

\appendix
\renewcommand{\theequation}{\Alph{section}\arabic{equation}}

\section*{Appendix}
\setcounter{equation}{0}

In this appendix present the double differential decay rate with respect to
the angle $\theta$ between $\ell^-$ and $\Lambda_b$, and dimensional
invariant mass of the leptons. Using the same convention and notation given
in \cite{R11118}, the double differential decay rate is given as:
\bea
\label{e111a1}
\frac{d\Gamma}{ds\,dx} = \frac{G^2\alpha^2 m_{\Lambda_b}}{16384 \pi^5}\vel
V_{tb}V_{ts}^\ast\ver^2 v \sqrt{\lambda(1,r,s)} \, \Big[ {\cal T}_0(s) +
{\cal T}_1(s) x + {\cal T}_2(s) x^2 \Big]~.
\eea
where $s = q^2/m_{\Lambda_b}^2$, $r = m_{\Lambda}^2/m_{\Lambda_b}^2$, 
$\lambda(1,r,s)=1+r^2+s^2-2r-2s-2rs$, 
$v=\sqrt{1-4 m_\ell^2/q^2}$ is the lepton 
velocity, $x=\cos\theta$.
The expressions for ${\cal T}_0(s)$, ${\cal T}_1(s)$ and 
${\cal T}_2(s)$ are:

\bea
\label{e111a2}
{\cal T}_0(s) \es 
%
32 m_\ell^2 m_{\Lambda_b}^4 s (1+r-s) \ga \vel D_3 \ver^2 +
\vel E_3 \ver^2 \dr \nnb \\
\ar 64 m_\ell^2 m_{\Lambda_b}^3 (1-r-s) \, \mbox{\rm Re} [D_1^\ast E_3 + D_3
E_1^\ast] \nnb \\
%
%
\ar 64 m_{\Lambda_b}^2 \sqrt{r} (6 m_\ell^2 - m_{\Lambda_b}^2 s)
{\rm Re} [D_1^\ast E_1] \nnb \\
\ar 64 m_\ell^2 m_{\Lambda_b}^3 \sqrt{r} 
\Big\{ 2 m_{\Lambda_b} s {\rm Re} [D_3^\ast E_3] + (1 - r + s) 
{\rm Re} [D_1^\ast D_3 + E_1^\ast E_3]\Big\} \nnb \\
%
%
\ar 32 m_{\Lambda_b}^2 (2 m_\ell^2 + m_{\Lambda_b}^2 s)
\Big\{ (1 - r + s) m_{\Lambda_b} \sqrt{r} \,
\mbox{\rm Re} [A_1^\ast A_2 + B_1^\ast B_2] \nnb \\
\ek m_{\Lambda_b} (1 - r - s) \, \mbox{\rm Re} [A_1^\ast B_2 + A_2^\ast B_1] - 
2 \sqrt{r} \Big( \mbox{\rm Re} [A_1^\ast B_1] + m_{\Lambda_b}^2 s \,
\mbox{\rm Re} [A_2^\ast B_2] \Big) \Big\} \nnb \\
\ar 8 m_{\Lambda_b}^2 \Big\{ 4 m_\ell^2 (1 + r - s) + 
m_{\Lambda_b}^2 \Big[(1-r)^2 - s^2 \Big] 
\Big\} \ga \vel A_1 \ver^2 +  \vel B_1 \ver^2 \dr \nnb \\ 
\ar 8 m_{\Lambda_b}^4 \Big\{ 4 m_\ell^2 \Big[ \lambda + 
(1 + r - s) s \Big] + 
m_{\Lambda_b}^2 s \Big[(1-r)^2 - s^2 \Big] 
\Big\} \ga \vel A_2 \ver^2 +  \vel B_2 \ver^2 \dr \nnb \\
\ek 8 m_{\Lambda_b}^2 \Big\{ 4 m_\ell^2 (1 + r - s) - 
m_{\Lambda_b}^2 \Big[(1-r)^2 - s^2 \Big] 
\Big\} \ga \vel D_1 \ver^2 +  \vel E_1 \ver^2 \dr \nnb \\
%
%
\ar 8 m_{\Lambda_b}^5 s v^2 \Big\{
- 8 m_{\Lambda_b} s \sqrt{r}\, \mbox{\rm Re} [D_2^\ast E_2] +
4 (1 - r + s) \sqrt{r} \, \mbox{\rm Re}[D_1^\ast D_2+E_1^\ast E_2]\nnb \\
\ek 4 (1 - r - s) \, \mbox{\rm Re}[D_1^\ast E_2+D_2^\ast E_1] +
m_{\Lambda_b} \Big[(1-r)^2 -s^2\Big]   
\ga \vel D_2 \ver^2 + \vel E_2 \ver^2\dr \Big\}\nnb
%
~,\\ \nnb \\
\label{e111a3}
{\cal T}_1(s) \es 
%
-32 m_{\Lambda_b}^4 s v \sqrt{\lambda}
\Big\{ \mbox{\rm Re}[A_1^\ast D_1] -  \mbox{\rm Re}[B_1^\ast E_1]
\nnb \\
\ar m_{\Lambda_b}
\mbox{\rm Re}[B_1^\ast D_2 - B_2^\ast D_1 +
A_2^\ast E_1 - A_1^\ast E_2] 
\nnb \\
%
%
\ek m_{\Lambda_b}^2 (1- r)\,
\mbox{\rm Re}[A_2^\ast D_2 - B_2^\ast E_2] \nnb \\
\ek m_{\Lambda_b} \sqrt{r} \, 
\mbox{\rm Re}[A_2^\ast D_1 + A_1^\ast D_2 - B_2^\ast E_1 - B_1^\ast E_2]
\Big\}~, \nnb
%
\\ \nnb \\
\label{e111a4}
{\cal T}_2(s) \es - 8 m_{\Lambda_b}^4 v^2 \lambda 
\Big\{ \vel A_1 \ver^2 + \vel B_1 \ver^2 + \vel D_1 \ver^2 
+ \vel E_1 \ver^2 \nnb \\
%
%
\ek m_{\Lambda_b}^2 s
\Big( \vel A_2 \ver^2 +
\vel B_2 \ver^2 + \vel D_2 \ver^2 + \vel E_2 \ver^2 \Big) \Big\}~. \nnb
\eea

\newpage

\section*{Figure captions}
{\bf Fig. (1)} The dependence of the differential branching ratio for the
$\Lambda_b \rar \Lambda \mu^+ \mu^-$ decay on $q^2$ for two different
parameter sets of the scenarios $S1$, $S2$. For a comparison we also present
the $SM$ result. \\ \\
{\bf Fig. (2)} The same as in Fig. (1), but for the forward--backward
asymmetry. \\ \\ 
{\bf Fig. (3)} The dependence of the asymmetry parameter
$\alpha_{\Lambda}$ on $q^2$ for two different parameter sets of the 
scenarios $S1$, $S2$ and $SM$. \\ \\
{\bf Fig. (4)} The same as in Fig. (3), but for the asymmetry parameter
$\alpha_\theta$.

\newpage

\begin{figure}
\vskip 1.5 cm
    \includegraphics{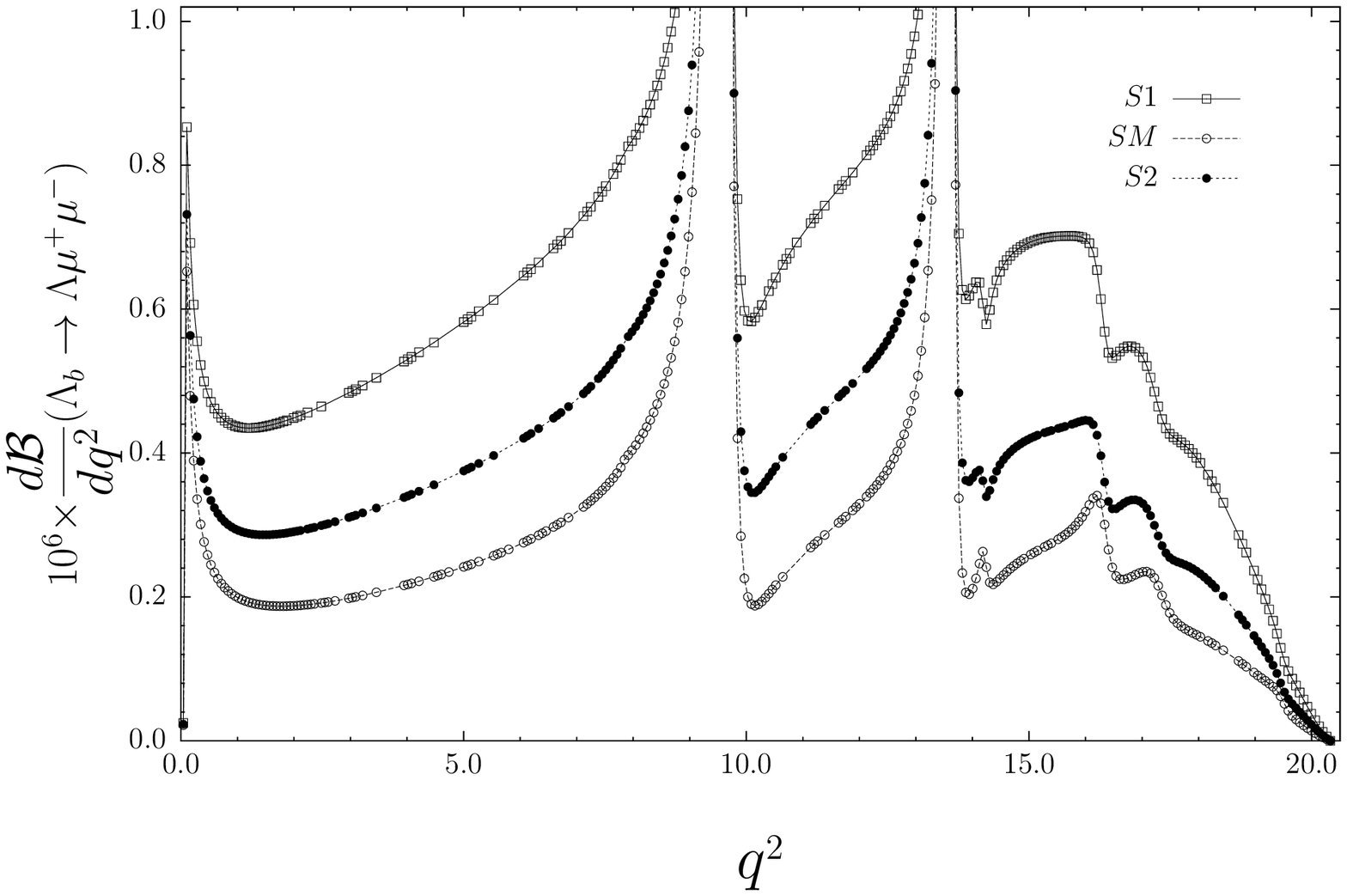}
\vskip 7.5cm
\caption{}
\end{figure}  

\begin{figure}   
\vskip 3.25 cm
    \includegraphics{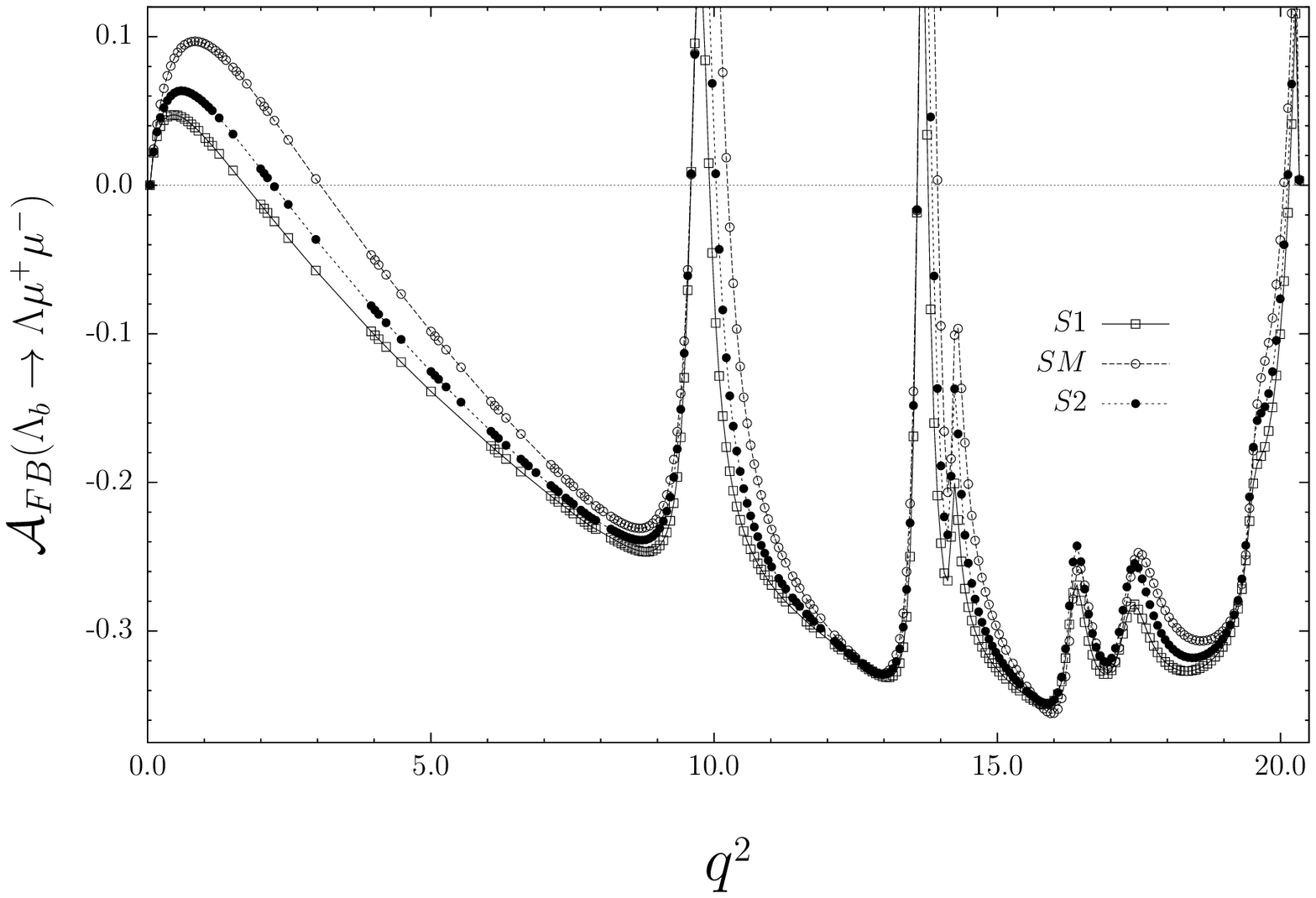}
\vskip 7.5 cm   
\caption{}
\end{figure}

\begin{figure}   
\vskip 1.5 cm
    \includegraphics{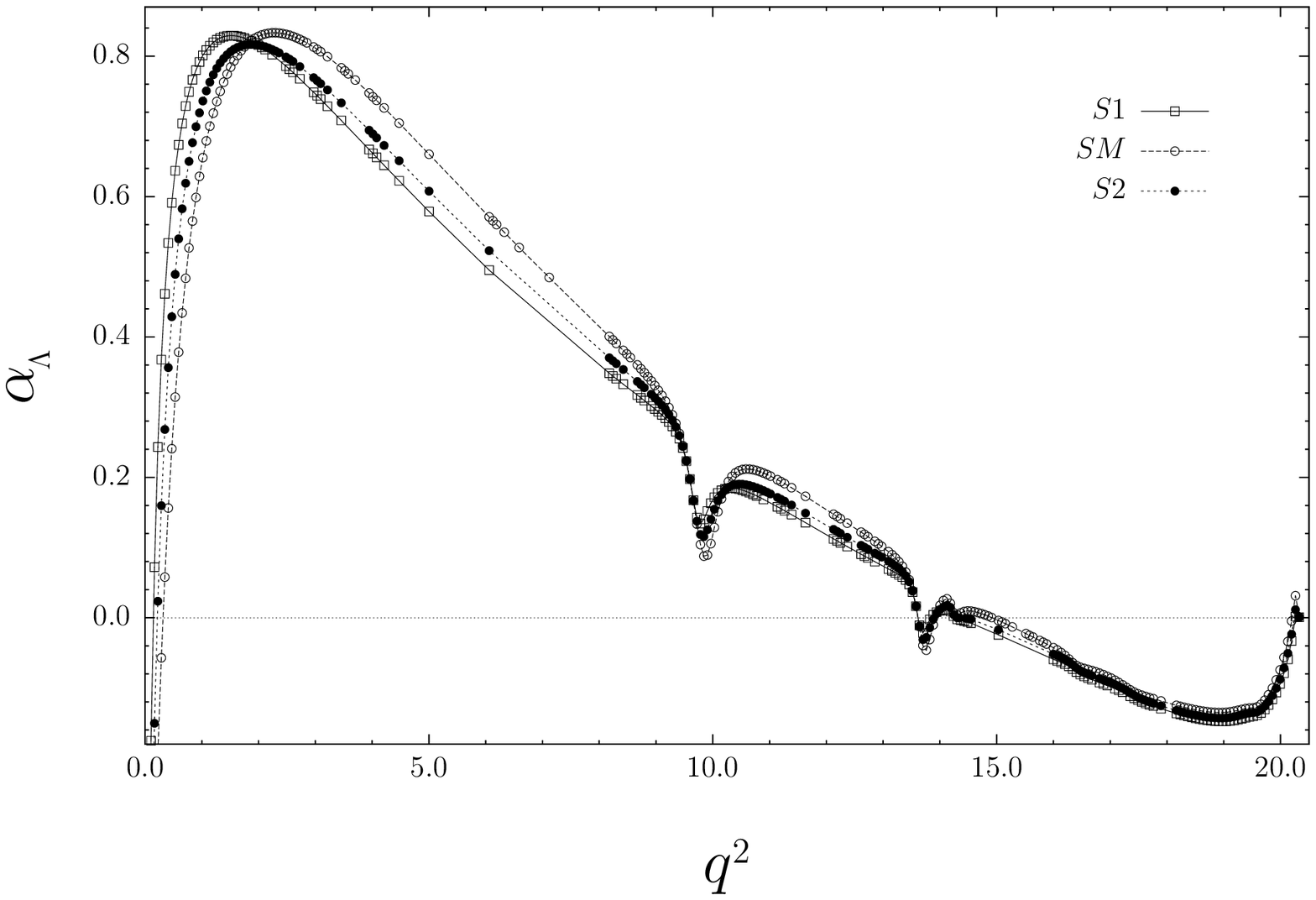}
\vskip 7.5cm
\caption{}
\end{figure}

\begin{figure}    
\vskip 3.25 cm
    \includegraphics{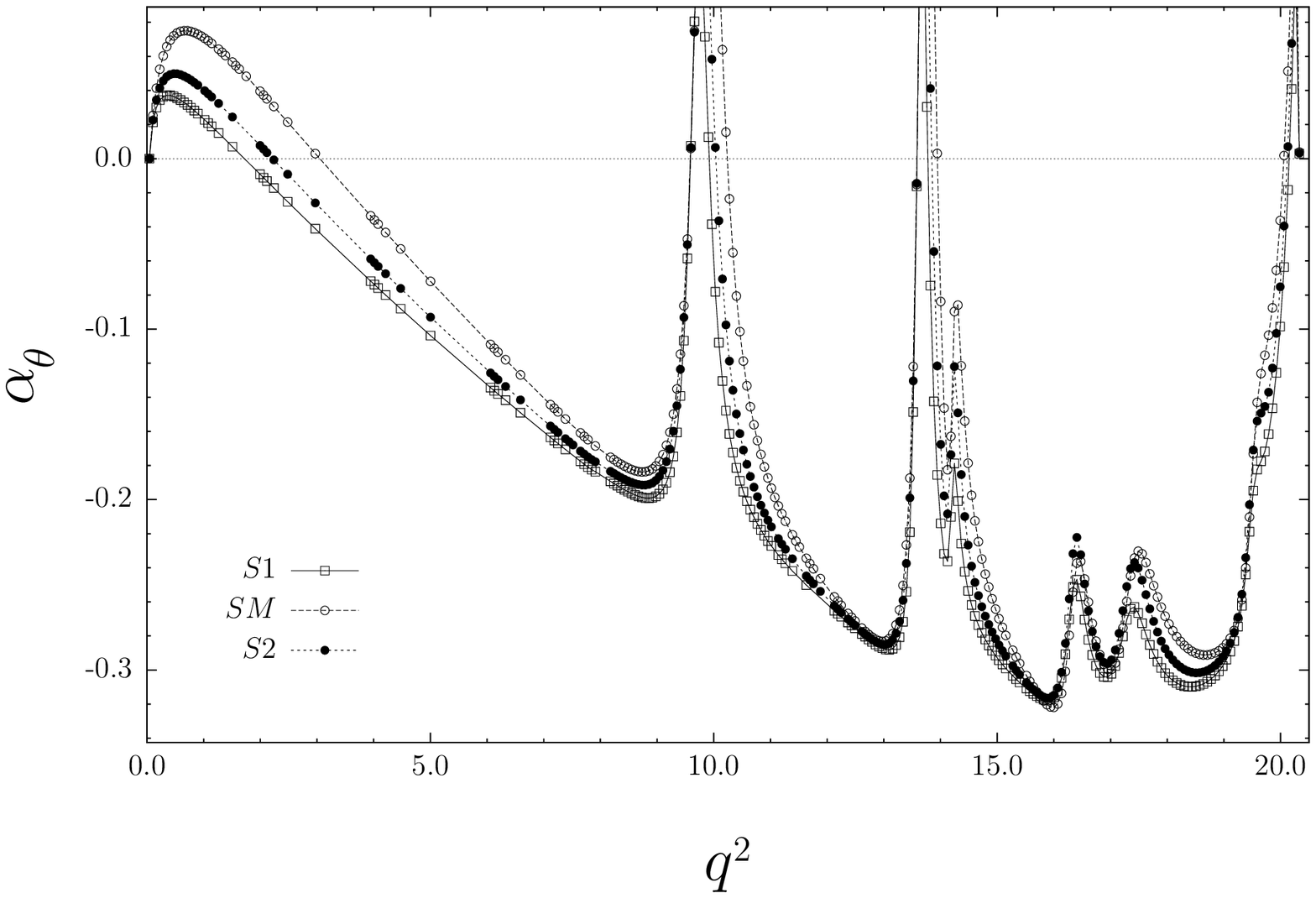}
\vskip 7.5 cm   
\caption{}
\end{figure}

\end{document}